\documentstyle[11pt,epsfig,amstex,amssymb]{article}                                           
\newcommand{\q}{$Q^2~$}

\newcommand{\f}{$F_2~$}                                            
\newcommand{\F}{$F_2(x,Q^2)~$}                                            
\newcommand{\beq}{\begin{equation}}                                            
\newcommand{\eeq}{\end{equation}}                            
\newcommand{\ee}{$e^+e^-~$}                 

\newcommand{\av}[1]{\mbox{$ \langle #1 \rangle $}} 
\begin{document}                                                               
\begin{titlepage}                                                              
\hfill
\hspace*{\fill}
\begin{flushleft}      
\noindent     {\tt DESY 96-143    \hfill    ISSN 0418-9833} \\    
{\tt August 1996} \\          
\end{flushleft}         
\vspace*{2.cm}                                                 
                
\begin{center}                                                                 
\begin{LARGE}                                                                  
{\bf  
An Observation on $F_2$ at Low $x$
}\\
\end{LARGE}                                                                    
\vspace{3cm}  
\begin{Large}                                                                 
{
A. De Roeck$^1$, E. A. De Wolf$^2$
}\\ 
\vspace{1cm}                                                   
$^1$ DESY, Notkestr.85, D-22607 Hamburg, Germany \\
$^2$ Universitaire Instelling Antwerpen, B-2610 Wilrijk, Belgium\\

\end{Large}
\end{center}                                                                   
\vspace*{5.cm}                          
\begin{center}                                       
{\bf Abstract}
\end{center}                                                                
\begin{quotation}                               
A simple  parametrisation of  H1 and ZEUS data at HERA is
given for the ranges in $x$ and \q of $ 10^{-4} - 5.10^{-2}$
and $5 - 250$ GeV$^2$, respectively.
This   empirical expression is based on a strikingly similar
dependence of the average charged particle 
multiplicity $\av{n}$ on the centre of mass system energy
$\sqrt{s}$ in \ee collisions on the one hand, and
the $x$ dependence of the proton 
structure function \f as measured at small $x$ on the other hand.
To the best of our knowledge, this similarity has not been noted before.
                                
\end{quotation}                                                                
\vfill                                                                         
\vspace{1cm}
\end{titlepage}                                                                
\newpage                                                                       
\vspace{2cm}                                                                   
\noindent                                                                      
One of the most successful tests of perturbative QCD is the quantitative
explanation of scaling violations, i.e. the \q dependence of 
the nucleon structure functions at fixed $x$-values. Here \q
and $x$ denote the usual deep-inelastic variables, the 
four-momentum squared and Bjorken-$x$.
Previous fixed target experiments measured the structure 
function \F for the region $x > 0.01$~\cite{NMC,BCDMS}. These data
 were, therefore,
sensitive to the valence content of the nucleon. 
The DGLAP evolution
equations\cite{DGLAP} describe the \q evolution of the structure 
functions in this region very well.
However, the data from the 
electron-proton collider HERA explore a new kinematic region. 
Values of Bjorken-$x$
 in the range
$10^{-5} < x <  10^{-2}$ for \q larger than 1 GeV$^2$, are reached.
In this region 
the valence contribution is expected to be negligible and  
\f to be driven by the gluon in the proton.

Recently new data on \f from the H1\cite{H1} and ZEUS\cite{ZEUS} experiments 
at HERA, based on the 1994 data taking period, have been published. The data
have reached a  level of precision  
of 3-5\% in a large region of the kinematical plane.
They show very clearly that \f rises 
strongly for decreasing $x$ for all \q values, and strong scaling
violations are observed in the new deep-inelastic region at low-$x$.
Originally it was thought that in the HERA region  $\ln 1/x$ terms, 
not included in the DGLAP resummation, could become important. 
However,
it has turned out that 
these evolution equations are still successful in describing the \q
dependence of the data~\cite{H1}. 

The rise of \f at small $x$
was predicted more than twenty years ago~\cite{Ruju} 
from the leading order renormalization
group equations of perturbative QCD. 
Ball and Forte recently pursued these ideas~\cite{Ball}
and proposed a way to demonstrate
that the low-$x$ data at HERA exhibit double asymptotic
scaling (DAS)
dominantly generated by  QCD radiation. They obtained an expression for
\F in the double asymptotic limit of low-$x$ and large $Q^2$.
The recent \F measurements of  H1 for 
$Q^2>5$ GeV$^2$ are broadly in agreement with  such a 
scaling behaviour. Hence, in this region
these data are expected to be sensitive to
the fundamental  QCD evolution dynamics, and  not to depend on unknown
(non-perturbative) starting distributions at sufficiently large $Q^2$
and small $x$.
This idea has also been exploited in the dynamically generated 
parton distributions\cite{GRV} which predicted, for the same reason, the 
rise of \f at small $x$ prior to data.
Qualitatively these results can be understood by viewing the deep inelastic 
collision at low-$x$ as the interaction of a virtual photon with partons
in a space-like parton cascade
 which stretches from
  $x$ of order one to $x << 1$, and thus covers a
   rapidity range $\propto \ln(1/x)$. For very small $x$, the rapidity
range is large and  
   a well-developed  cascade can be formed.
In the leading-log approximation this leads to an 
expression~\cite{DGLAP} for \f
\begin{equation}
F_2 \sim \exp{\sqrt{\frac{16 N_c}{b} \ln( 1/x)\ln\ln Q^2}}.\label{f2ass}
\end{equation}
Here $b$ is the
leading coefficient in the $\beta$-function for the expansion of
$\alpha_s$, namely $b=11-2n_f/3$       
with $n_f$ the number of flavours; $N_c$ is the number of colours.

Another cornerstone of the success of perturbative QCD are the calculations
and predictions for particle production in time-like parton cascades
 in \ee collisions, based on the Modified Leading Log 
Approximation (MLLA) evolution equations 
and the assumption of  Local Parton Hadron Duality
(LPHD)\cite{ochs,khoze}.
In this approximation, the average 
parton multiplicity of \ee collisions as function
of the center of mass system energy $\sqrt{s}$ (CMS) is given by:
\begin{equation}
<n_p> \sim \Gamma(B)(\frac{z}{2})^{1-B}I_{B+1}(z),\label{bessel}
\end{equation}
with $z= \sqrt{(16 N_c/b) \cdot \ln (\sqrt{s}/2Q_1)}$.
The function $I_{B+1}(z)$ is a 
Bessel function of order $B+1$, with, for four 
flavours, $B=(11 + 2n_f/27)/b=1.355$. Here 
$\Gamma$ is  the Gamma function.
The parameter $Q_1$ is the $p_t$ cutoff of the 
partons in the shower and found to be in the
range
of 250-290 MeV~\cite{ochs} from fits to the data.
Eqn.~(\ref{bessel}) gives a very good description of
the averaged charged hadron  multiplicity in \ee collisions  for CMS energies
in the range $\sqrt{s} = 3-130$ GeV\cite{ochs}.

\begin{figure}[htbp] 
{\epsfig{file=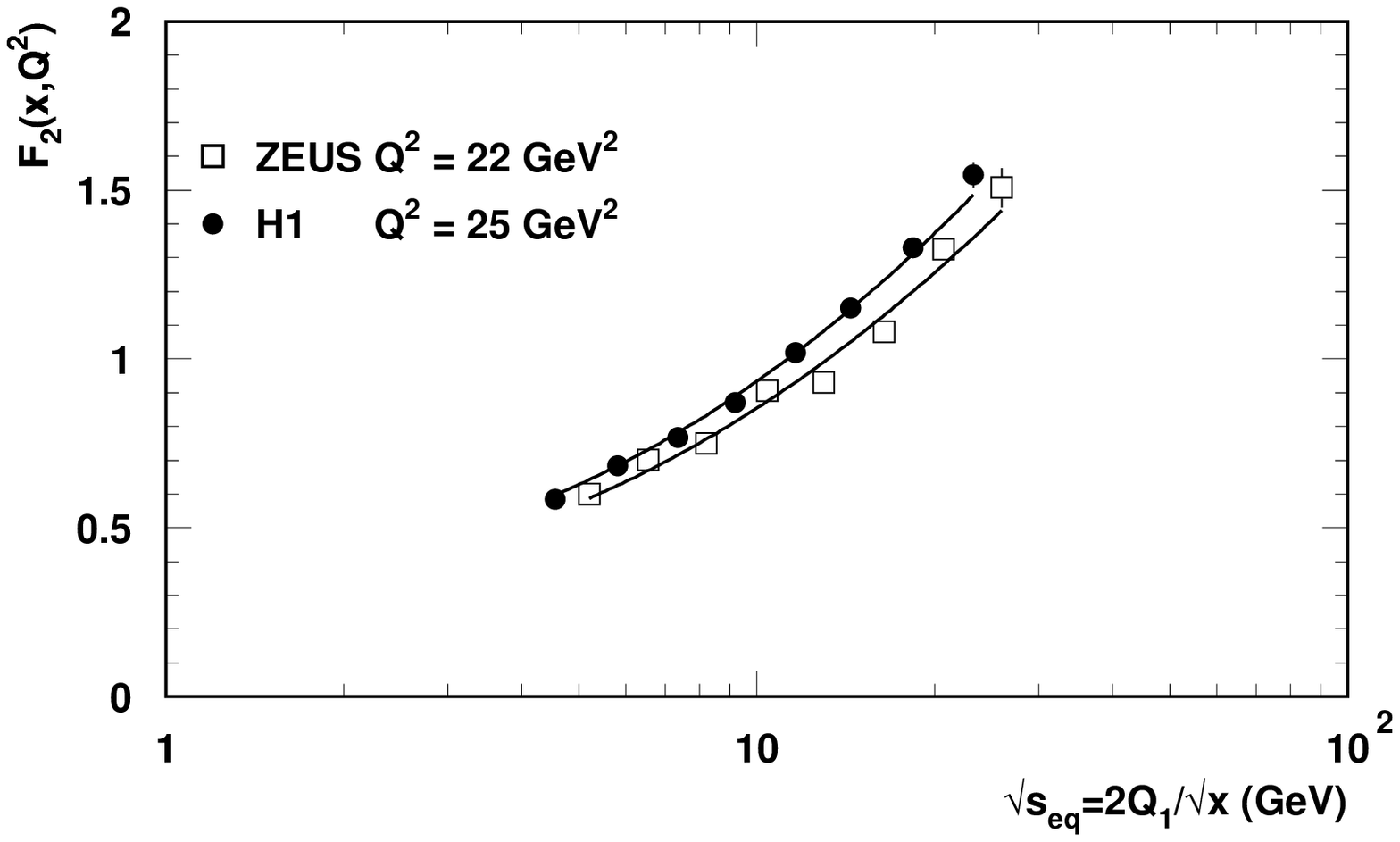,width=15cm,bbllx=50pt,bblly=150pt,bburx=500pt,bbury=970pt,angle=-90}}
\caption[]{ \label{f0}
Comparison of \ee data on average charged 
particle multiplicities
versus $\sqrt{s}$ and the HERA 
low-$x$ \f data versus $2Q_1/\sqrt{x}$, with $Q_1 = 270$
MeV, for $Q^2 = 22$ GeV$^2$ (ZEUS) and 
25 GeV$^2$ (H1). The \ee multiplicity data (solid lines) are represented
by curves resulting from a phenomenological fit through the data
\protect \cite{OPAL}.
The curves are normalized to the \f data for each $Q^2$ bin separately.}
\end{figure}

Expression (\ref{bessel}) can be approximated at large $z$ by 
\begin{equation}
<n_p> \sim \exp{\sqrt{\frac{16 N_c}{b} \ln \sqrt{s}/2Q_1}} \label{multip}.
\end{equation}
Comparing expressions (\ref{f2ass}) and (\ref{multip}), 
one notices an intriguing
similarity.
For fixed $Q^2$ they have a similar functional dependence on $1/x$ and
$s/4Q_1^2$, respectively.
The connection $s \rightarrow 1/x$      emerges naturally in
Regge-inspired phenomenology, see e.g.\cite{Landshoff}.
In Fig.~\ref{f0} we compare the \ee data on average charged 
particle multiplicities
versus $\sqrt{s}$ and the HERA 
low-$x$ \f data versus $2Q_1/\sqrt{x}$, with $Q_1 = 270$
MeV, as determined in \cite{ochs}. The \ee multiplicity data are represented
by curves resulting from a phenomenological fit through the data as
derived by OPAL\cite{OPAL}. 
The curves are normalized to the \f data for each $Q^2$ bin separately.
It shows that at small $x$ the evolution  of \f with  $1/x$ and the 
dependence of average charged particle multiplicity in \ee collisions
on  $\sqrt{s}$  are indeed quite similar as suggested by the expressions 
above.

This simple observation 
led us to study fits of the full expression (\ref{bessel})
 to the new low-$x$ measurements of   \f
at HERA, with the change $s/4Q_1^2 \rightarrow 1/x$.
The \q dependence (absent in \ee) was assumed 
to be given by a  slowly varying function of $Q^2$ of the form 
 $(\ln \alpha_s(Q^2_0)/\alpha_s(Q^2))^{\delta}$, with $\delta$ taken 
to be a constant.
The final expression fitted to the data is
\begin{equation}
F_2(x,Q^2)= C(Q^2)\Gamma(B)(\frac{z}{2})^{1-B}I_{B+1}(z),
\label{f2bessel}
\end{equation}
and
\begin{equation}
z=\sqrt{\frac{16N_c}{b}\ln \sqrt{1/x}\cdot (\ln
\alpha_s(Q^2_0)/\alpha_s(Q^2))^{\delta}};\label{finz}
\end{equation}                        
where  $Q_0$ is taken to be 1 GeV and the two-loop 
expression for $\alpha_s$ is used with $\Lambda_{QCD}^{(4)} = 263$ 
MeV~\cite{virchaux}.
Note that the normalization factor
$C(Q^2)$ and power $\delta$  are
{\sl the only fit parameters} at any fixed $Q^2$.

\begin{figure}[htbp] 
{\epsfig{file=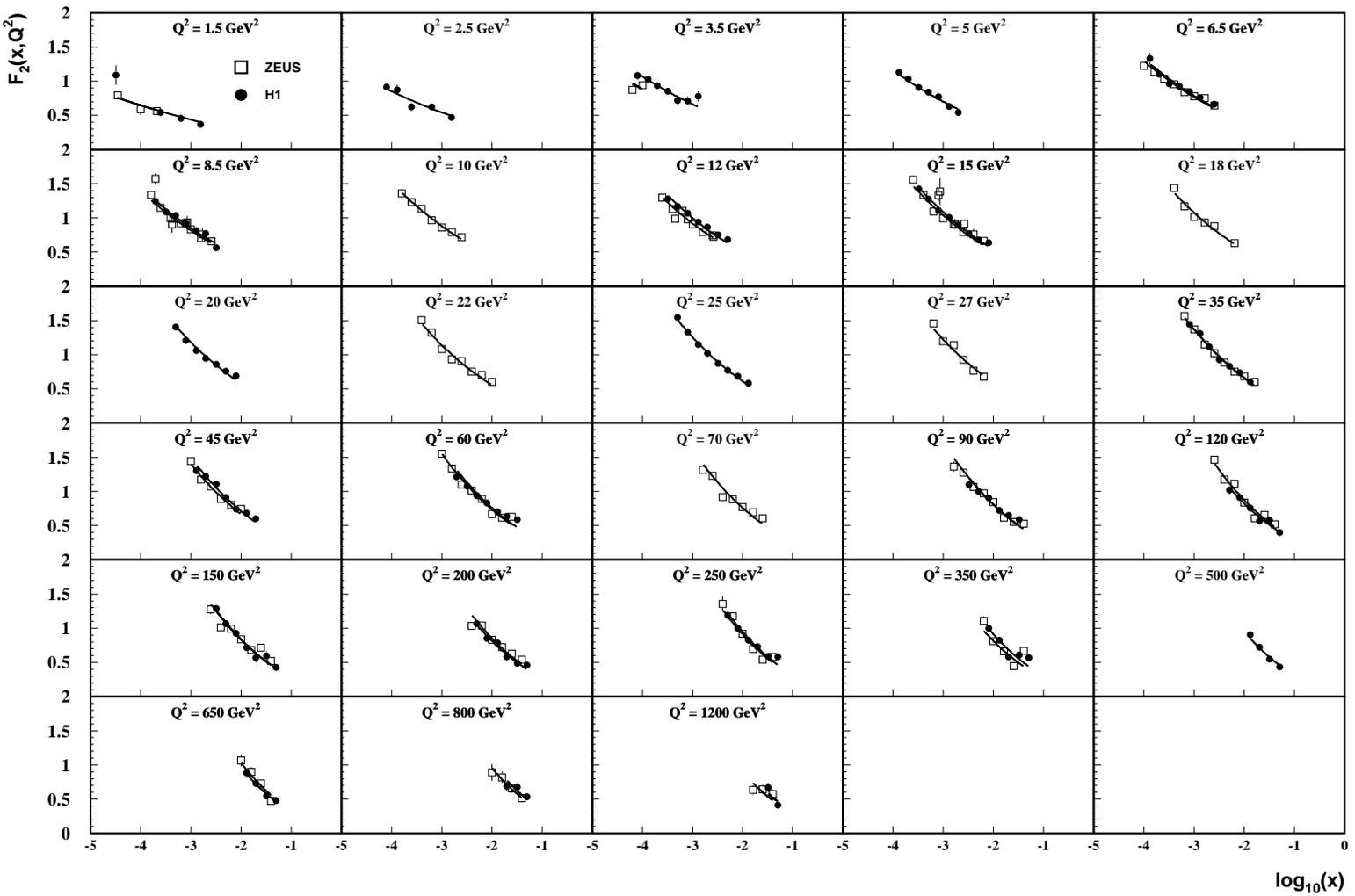,width=15cm,bbllx=0pt,bblly=0pt,bburx=540pt,bbury=802pt}}
\caption[]{ \label{f1}
The proton structure function $F_2$ measured by the H1 and ZEUS experiments
at HERA together with fits of eqn.~(4)
through both data sets separately. The normalization constant $C(Q^2)$ was 
fitted separately for each $Q^2$ value.}
\end{figure}

\begin{figure}
{\epsfig{file=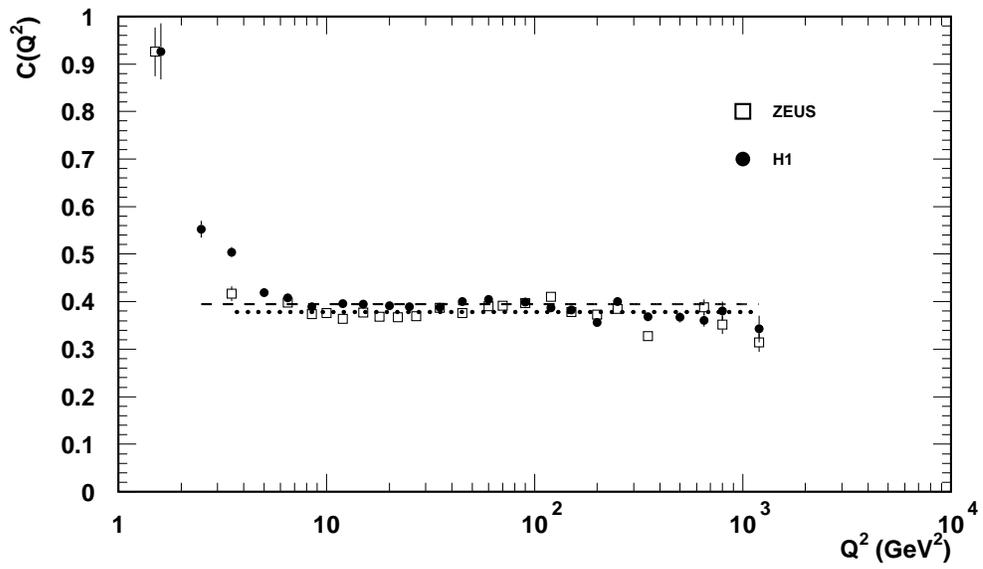,width=15cm,bbllx=0pt,bblly=120pt,bburx=540pt,bbury=802pt,angle=-90}}
\caption[]{ \label{f2}
The normalization constant $C$ of eqn.~(4) as function of $Q^2$.}
\end{figure}

The result is shown in Fig.~\ref{f1}, where a fit 
is made in  each bin of \q  on the H1  and ZEUS                  
data, separately. The expression (\ref{f2bessel}) describes the data
well over the whole kinematic region, except at large $y
=Q^2/xs$ values, where the contribution of valence quarks is expected to
become important. The  difference
in the results        obtained using the H1 or ZEUS
data can 
hardly be distinguished. 
We find that the data are best reproduced for $\delta \sim 0.7$ (see below),
definitely below the value $\delta = 1$ derived from the asymptotic form 
in perturbation theory\cite{Ruju,Ball}. 
The result for the normalization $C$ is shown 
in Fig.~\ref{f2} as function of $Q^2$.
 In the range $5<Q^2<250$ GeV$^2$, $C$ is essentially constant with a value of 
about 0.38. For lower $Q^2$, a clear breaking of this regularity is observed,
and  hints that additional contributions to \f become important.

\begin{figure}
{\epsfig{file=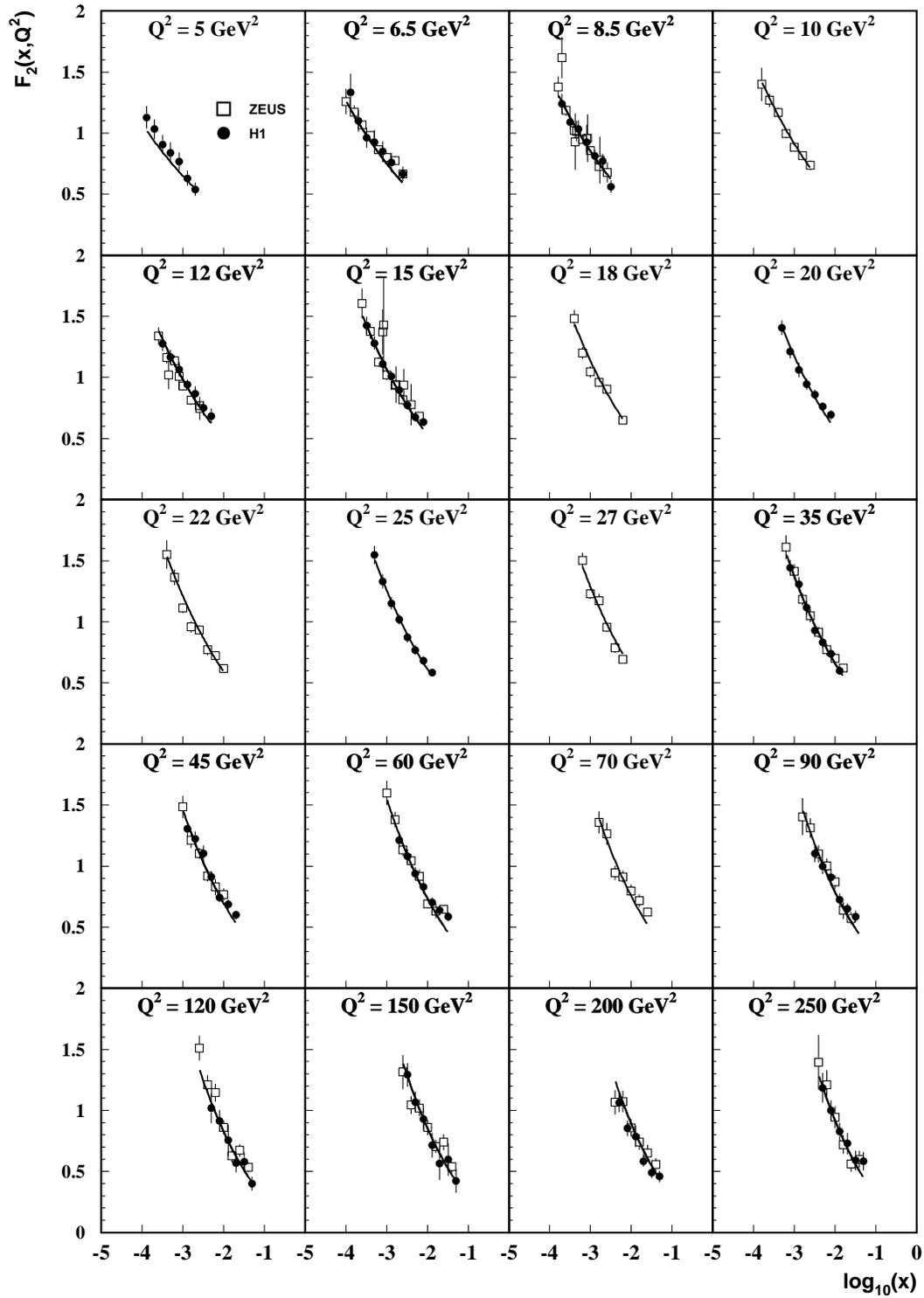,width=15cm,bbllx=0pt,bblly=0pt,bburx=540pt,bbury=802pt}}
\caption[]{ \label{f3}
The proton structure function $F_2$ in the range $5<Q^2<250$
GeV$^2$, $x<0.05$ and
$y>0.02$ by the H1 and ZEUS experiments together with a fit of eqn.~(4), 
for which
the normalization constant $C$ is taken to be independent of $Q^2$.}
\end{figure}

Encouraged by the results shown in 
Figs.~\ref{f1} and \ref{f2}  we 
perform a combined fit of the H1 and ZEUS
data to  eqn.~(\ref{f2bessel})
with $C(Q^2)=C_0 $  constant over the whole \q range, in the region 
$5<Q^2< 250$ GeV$^2$, $x< 0.05$, $y>0.02$. The latter two conditions
are imposed to avoid the valence quark region.
The result is shown in Fig.~\ref{f3}. 
The fit has $\chi^2/NDF = 265/231$, using the full errors.
The relative normalization of the H1 and
ZEUS data was left free. The normalization factors found are 0.99 and 1.025
for H1 and ZEUS respectively, well within the quoted normalization
uncertainties~\cite{H1,ZEUS}.
The statistical errors on the 
fit parameters are from a fit with the statistical errors of the data
only. Using the full error matrix of H1 and/or ZEUS each of the 
measured quantities entering the $F_2$ analysis is varied in turn.
For the two fit parameters we find $C_0=0.389\pm 0.005(stat) 
\pm 0.012(syst)$ and 
$\delta= 0.708\pm 0.007 (stat) \pm 0.028(syst)$.                     
From the fits to data of the individual experiments we find for H1:
 $C_0=0.385\pm 0.007(stat) \pm 0.020(syst)$ and 
$\delta= 0.683\pm 0.010 (stat) \pm 0.055(syst)$
($\chi^2/NDF = 76/97$); 
for ZEUS:
  $C_0=0.384\pm 0.007(stat) \pm 0.009(syst)$ and 
$\delta= 0.723\pm 0.010 (stat) \pm 0.025(syst)$ 
($\chi^2/NDF = 186/134$).                     
 A point by point analysis shows that the region $y< 0.04$ is
responsible for a substantial
 contribution to the $\chi^2$ for the ZEUS data.

With two free parameters only: the normalization $C_0$ and $\delta$, we are 
able to account for the $x$ and \q dependence of \f starting from a 
parametrisation which successfully describes the energy dependence of the 
mean charged multiplicity in \ee annihilation, provided $s$ is identified 
with $1/x$. We also note that according to eqn.~(\ref{f2bessel}) \f grows  
slower than any power of $1/x$ but faster than any power of $\ln 1/x$. In
particular, for most of the regions in \q shown in Fig.~\ref{f3}, 
the \f data indeed increase faster than $\ln 1/x$, contrary to the 
claims in~\cite{haidt}.

Fig.~4 shows
$\lambda = d \ln F_2(x,Q^2)/d\ln (1/x)$ 
calculated from (\ref{f2bessel})
 for a number of              $x$ values. 
 A rise of $\lambda$ with \q is observed.
Note that its value   depends on the $x$-region: 
 $\lambda$ increases with increasing $x$.                                        
The growth of $\lambda$ with $Q^2$ is often considered to be
indicative of a transition from a region of
``soft'' pomeron exchange ($\lambda\sim0.1$) at low $Q^2$ to a
regime of ``hard'' pomeron exchange ($\lambda\sim0.3-0.4$)
at high $Q^2$.
This  argument is based on measurements  of $d \ln F_2(x,Q^2)/d\ln (1/x)$
which  cover, however,  different ranges in $x$ as $Q^2$ changes.
Fig.~4 demonstrates that the so-called ``soft'' to ``hard''
transition is much less spectacular when $x$ is kept fixed.
In addition, we note that the slopes at a given $Q^2$ are
{\em larger} at large $x$ than at small $x$.
This runs contrary to the often expressed opinion that the small $x$
region in deep-inelastic scattering probes the ``hard'' pomeron.

In first instance we regard  eqn.~(\ref{f2bessel})  
as   a compact parametrisation of 
the \f data at small $x$,  where the dynamics of the 
\f evolution is expected to be dominated by  gluons. Since it 
is based on  a result of the MLLA evolution equations, which include 
coherence,
it is well adapted to           
be used e.g. as an ansatz for  starting distributions in QCD fits
of proton structure data.  

However, it is tempting to speculate that the 
similarity observed here  is more than just a 
mathematical coincidence.  It indeed suggests that, at least qualitatively,
the evolution of the structure function at low-$x$ can be attributed to
the development of an unhindered 
 QCD parton shower in  ``free'' phase space quite similar to that in \ee. 
For \f this also follows 
essentially from the observation of DAS and the success of the dynamically
generated GRV parton distributions. 
Whether a more profound explanation for the empirical regularity reported 
here exists, remains an interesting open question.

\section*{Summary}
A striking similarity between the rise with energy ($\sqrt{s}$) of the 
charged particle multiplicity in \ee and 
the rise of \f at HERA is observed.
To the best of our knowledge, this similarity has not been noted before.
For $Q^2 \geq 5$ GeV$^2$ and $10^{-4}<x<0.05$,
the phenomenologically successful 
MLLA expression for the average multiplicity in \ee
collisions, with the transformation $ s \rightarrow 1/x$, and adding
a QCD inspired 
 \q dependence, describes the HERA data on \f at small $x$ very well.
The result suggests  that both deep inelastic small-$x$ scattering 
and \ee annihilation can  be adequately described 
by angular ordered QCD radiation in an essentially free phase space.

\section*{Acknowledgement}
We thank V. Khoze, L. Lipatov and W. Ochs for stimulating discussions.

\begin{figure}
{\epsfig{file=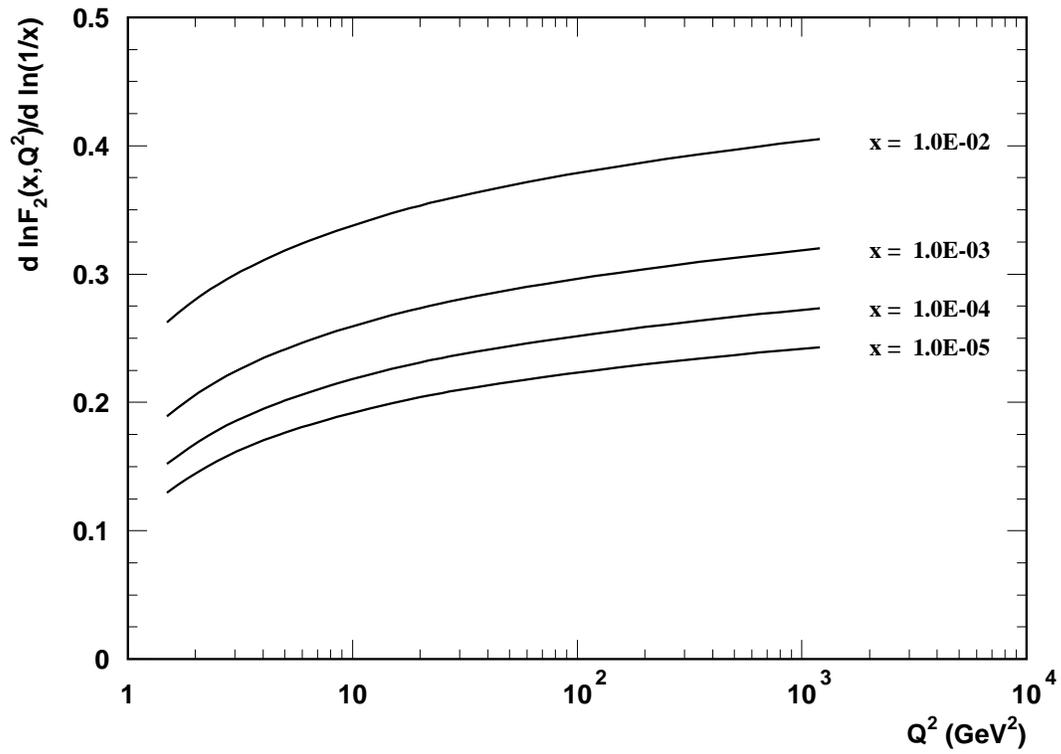,height=15cm,bbllx=0pt,bblly=200pt,bburx=300pt,bbury=700pt}}
\caption[]{ \label{f4}
The  slope parameter $\lambda = d \ln F_2(x,Q^2)/d\ln (1/x)$ as function
of $Q^2$,
derived from the final parametrisation for 
several $x$ values.}
\end{figure}
\end{document}